\documentclass[aps,prl,superscriptaddress,reprint,amsmath,amssymb]{revtex4-2}

\makeatletter

\def\@bibdata{}
\def\@bibtail{}
\def\bibliography#1{\gdef\@bibdata{#1}}
\def\bibliographystyle#1{}
\makeatother

\usepackage{ragged2e}
\usepackage{mathptmx}
\usepackage{upgreek}

\usepackage{graphicx}
\usepackage[colorlinks,bookmarksopen,bookmarksnumbered,citecolor=blue, linkcolor=blue, urlcolor=blue]{hyperref}

\usepackage{soul,color,xcolor}      
\definecolor{myColor}{rgb}{0,0,0}        
\makeatletter
\newcommand*{\revise}{\@ifnextchar\bgroup{\revise@}{\color{myColor}}}
\newcommand*{\revise@}[1]{{\textcolor{myColor}{#1}}}
\makeatother

\begin{document}
	\title{Optical M{\"o}bius Snails}
	
\author{Qiang Wang}
\email[]{These authors contributed equally to this work.}
\affiliation{Department of Physics, College of Science, Shantou University, Shantou 515063, China}

\author{Pedro A. Quinto-Su}
\email[]{These authors contributed equally to this work.}
\affiliation{Instituto de Ciencias Nucleares, Universidad Nacional Aut{\'o} noma de {\'e}xico, Apartado Postal 70-543, 04510, Cd. Mx., M{\'e}xico}
\affiliation{Centre for Disruptive Photonic Technologies, School of Physical and Mathematical Sciences, Nanyang Technological University, Singapore 637371, Singapore}

\author{Xi Xie}
\affiliation{Centre for Disruptive Photonic Technologies, School of Physical and Mathematical Sciences, Nanyang Technological University, Singapore 637371, Singapore}

\author{Xiangsheng Xie}
\affiliation{Department of Physics, College of Science, Shantou University, Shantou 515063, China}

\author{Chenghou Tu}
\email{Corresponding author.\\tuchenghou@nankai.edu.cn}
\affiliation{School of Physics and Key Laboratory of Weak Light Nonlinear Photonics, Nankai University, Tianjin 300071, China}

\author{Yijie Shen}
\email{Corresponding author.\\yijie.shen@ntu.edu.sg}
\affiliation{Centre for Disruptive Photonic Technologies, School of Physical and Mathematical Sciences, Nanyang Technological University, Singapore 637371, Singapore}

	\begin{abstract}
The exploration of topological states has emerged as a vibrant frontier across diverse physical systems. Among these, the M{\"o}bius strip stands out as a canonical structure that has recently inspired intriguing optical physics. In this work, we extend this concept into the high-dimensional realm by \revise{using} an optical analogue—the optical M{\"o}bius snail.  Specifically, employing a skyrmionic beam as the input field, tight focusing yields a continuous family of Möbius strips that \revise{exists in} three‑dimensional space, giving rise to a snail‑like polarization topological texture. Furthermore, we demonstrate full-field polarization tomography and achieve \revise{switchable control} of this optical snail texture by tuning the Skyrme number of the incident skyrmionic beam. Our findings establish a direct bridge between low- and high-dimensional M{\"o}bius geometries, opening a new platform for exploring complex spatial topology and tailored light-matter interactions.

	\end{abstract}
	
	\maketitle

From abstract mathematics to the frontiers of photonics, topology has profoundly reshaped our understanding and control of light through its implementation in structured light—from optical vortices carrying orbital angular momentum \cite{Light-8-90-2019} and polarization singularities \cite{Light-12-238-2023,APL-6-040901-2021}, to the vectorial nature of the structured light \cite{Light-11-205-2022,NP-15-253-2021}. These topological constructs enable on-demand engineering of diverse states, such as phase dislocations \cite{Light-8-90-2019} and C-point singularities \cite{APL-6-040901-2021}. Among the most striking examples are optical skyrmions—topologically protected quasiparticles originally discovered in magnetic systems. In photonics, skyrmions emerge from intricate spatial textures of the electromagnetic field \cite{NP-18-15-2024,PRL-129-267401-2022,AOP-17-295-2025}, including evanescent fields at metal surfaces \cite{Science-361-993-2018,NC-13-8-2022,OE-28-10320-2020}, and tightly focused beams carrying orbital angular momentum \cite{NP-15-650-2019}. They also manifest in the spin angular momentum density \cite{NP-15-650-2019,PRL-127-237403-2021,ACS-8-2567-2021,Lpr-19-2400327-2025,AP-7-016009-2025}, the Poynting momentum vector \cite{PRL-133-073802-2024,Nano-14-2211-2025}, and the Stokes vector fields on the Poincaré sphere \cite{OL-46-3737-2021,PR-11-2042-2023,Lpr-17-23000155-2023,PRL-134-083802-2025}. \revise{The inherent stability of these nontrivial skyrmion topologies, which are robust against small perturbations, makes them highly promising for applications such as high-dimensional optical communications and robust topological data storage, and has been confirmed and reported in numerous studies} \cite{NE-05944-2026}.

Optical Möbius strips are a typical kind of polarization topology, where a singularity‑driven structure organizes the polarization ellipse orientation into configurations of twisted ribbons (Möbius strips) along closed in‑plane trajectories \cite{Science-347-964-2015,OE-27-11516-2019,OE-27-29685-2019,OE-29-25535-2021,OL-48-4420-2023}. The first experimental observation was realized by tightly focusing Poincaré beams, revealing that the major axis of the polarization ellipse traces a Möbius strip around a C‑point, with the number of half‑twists determined by the topological charge of the input beam \cite{Science-347-964-2015}. Beyond these single‑twist configurations, multi‑twist polarization ribbons have also been realized in highly confined optical fields, where the polarization ellipse major axis undergoes multiple half‑twist along a closed loop, giving rise to higher‑order Möbius topologies \cite{SR-7-13653-2017,NJP-2019-21-053020}. The Möbius topologies were also observed in the focal field of a tightly focused linearly polarized beam \cite{PRL-117-013601-2016} and in ultrafast twisted ribbons formed by the instantaneous electric field vector \cite{Optica-7-1228-2020}. In parallel, theoretical and numerical studies revealed that optical polarization Möbius strips can also arise around C lines in the scattering from high-index dielectric nanoparticles, making all-dielectric scatterers as a promising platform for topologically protected polarization singularities \cite{ACS-4-1159-2017}. \revise{These results demonstrate a powerful paradigm for controlling the vectorial topology of light; however, such Möbius topologies have so far been exclusively confined to two‑dimensional (2D) planar cross‑sections: a single closed loop in the transverse plane yields a twisted ribbon, but the possibility of extending this concept into a continuous three‑dimensional (3D) volume has remained unexplored.}  

Extending the study into the high-dimensional space of light fields reveals even richer topological architectures~\cite{Light-11-205-2022,NP-15-253-2021}. For example, 3D topological states can assume diverse forms, including dark knots and braided optical lines \cite{NP-6-118-2010,NC-11-5119-2020,NP-14-1079-2018}, particle-like 3D hopfion textures~\cite{NC-12-6785-2021,PRL-131-263801-2023,AP-5-015001-2023}, photonic torons delivering phase transitions among many 3D topologies~\cite{PRL-135-063802-2025}, Shankar skyrmions with Hopf mapping to SO(3)~\cite{PRL-135-233803-2025}, as well as emerging 4D topologies with new dimensions for studying light \cite{SA-9-eadh0369-2023,PRL-134-123805-2025}. These advances naturally raise a fundamental question: Can a high-dimensional counterpart of the M{\"o}bius strip be created in light, extending its topological properties across a continuous volumetric region?

In this article, we bridge this gap by introducing and experimentally demonstrating the \textit{optical M{\"o}bius snails}, a 3D version beyond the planar confinement of the prior optical M{\"o}bius strips. Through tailored skyrmionic input beams, we design a family of 3D loops where the spatially varying orientation of the polarization ellipse's major axis forms a globally linked, coiled structure. We experimentally reconstruct this structure via full-field tomography under tight focusing. Furthermore, we achieve dynamic control over its topology by modulating the Skyrme number of the incident field, enabling transitions between different types of M{\"o}bius snails. This work establishes a new class of 3D topological structures in light, enriching topological photonics and opening avenues for applications in topological sensing, optical encryption, and higher-dimensional light-matter interactions.

\begin{figure}[!t]
	\centering
	\includegraphics [width=0.48\textwidth]{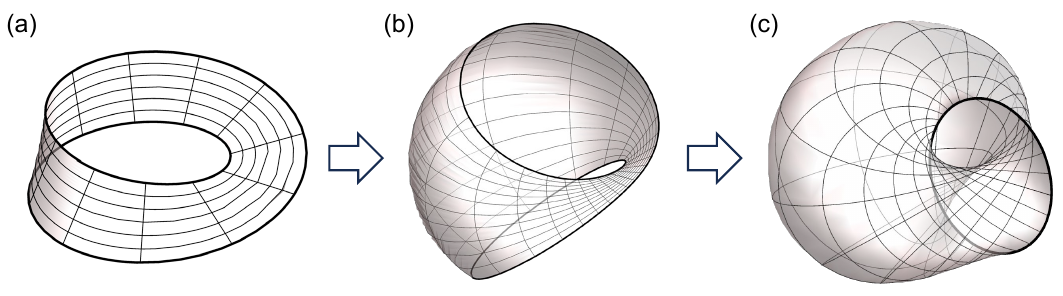}
	\caption{From the M{\"o}bius strip to the M{\"o}bius snail. (a) A standard M{\"o}bius strip. (b) Transitional geometry during the deformation of the M{\"o}bius strip into a M{\"o}bius snail. (c) A fully developed 3D M{\"o}bius snail, obtained by continuously shaping the M{\"o}bius strip so that its single boundary becomes a perfect circle in $\mathbb{R}^3$. In all three cases, the surface remains non‑orientable and possesses only one boundary curve.}
	\label{fig1}
\end{figure} 

In the diverse landscape of topology and geometry, the M{\"o}bius strip stands as a symbolic icon of one-sided surfaces. A particularly notable variant is the \textit{Sudanese M{\"o}bius strip}--a fascinating mathematical object in which the single boundary edge closes into a perfect circle \cite{AM-92-335-1970,SIG-1984}, discovered by Blaine Lawson and named in honor of topologists Sue Goodman and Daniel Asimov (Sue‑dan‑ese). Owing to its characteristic coiled geometry reminiscent of a snail's shell, it is also poetically referred to as the "M{\"o}bius snail". Mathematically, it can be viewed as a M{\"o}bius strip embedded in the three‑sphere $\mathbb{S}^3$, which is then mapped into ordinary 3D space $\mathbb{R}^3$ via stereographic projection \cite{arXiv-2012,PP-32-102503-2025}. A standard M{\"o}bius strip is formed by joining the ends of a rectangular strip after a half‑twist--resulting in a one‑sided surface with a single boundary, as illustrated in Fig. \ref{fig1}(a)--the M{\"o}bius snail generalizes this concept. By allowing the single edge of the M{\"o}bius strip to extend and close upon itself in a specific manner, one obtains a M{\"o}bius snail: a 3D topological geometry whose boundary forms a perfect circle in $\mathbb{R}^3$ (see Supporting Information for details). Figs. \ref{fig1}(b) and \ref{fig1}(c) show an intermediate stage in this geometric evolution and the fully formed 3D M{\"o}bius snail.

Here we realize optical M{\"o}bius snails by tightly focusing a laser beam that carries a Stokes‑skyrmion polarization texture, which can be generated through the coherent superposition of two circularly polarized Laguerre–Gaussian (LG) beams with opposite handedness \cite{OL-46-3737-2021,PRL-134-083802-2025}. Specifically, the transverse electric field of the input beam can be written as
\begin{equation} \label{eq1}
	\mathbf{E}_0 = {\rm LG}_{0,l_1} |R\rangle + {\rm LG}_{0,l_2} |L\rangle
\end{equation}

\revise{Where $|R\rangle$ and $|L\rangle$ are states of right- and left-handed circular polarizations, which form an orthonormal basis satisfying $\langle R | R\rangle = \langle L | L\rangle = 1$ and $\langle R | L\rangle = 0$.} ${\rm LG}_{p,l}$ represents an LG mode with radial and azimuthal indices ($p$, $l$). \revise{Here the amplitude ratio between the two circularly polarized components is taken as 1:1, and there is no additional phase offset between them.} The Stokes skyrmion texture is described by the normalized Stokes parameters $\mathbf{s}=(s_1,s_2,s_3)$, and the corresponding Skyrme number is given by $N=(1/4\pi) \iint \mathbf{s}\cdot (\partial_x \mathbf{s} \times \partial_y \mathbf{s}) dxdy$, \revise{with $2\psi={\rm atan2}(s_2/s_1)$, and $\rm atan2()$ returns values in the interval $[-\pi,\pi]$. Where the integration domain is the entire transverse plane.}

\begin{figure}[!t]
	\centering
	\includegraphics [width=0.48\textwidth]{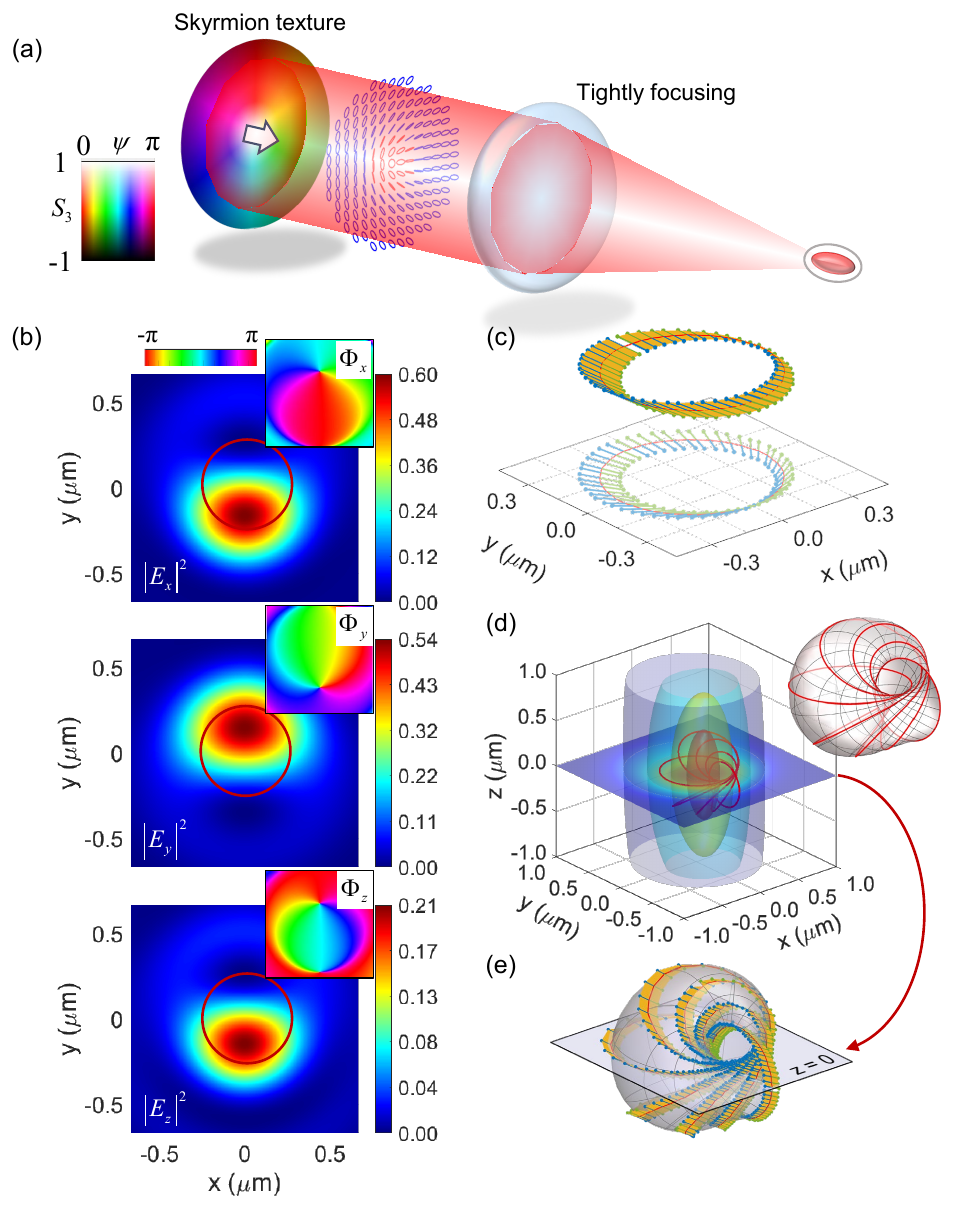}
	\caption{Simulated results for the optical M{\"o}bius strip and M{\"o}bius snail structures for a tightly focused skyrmionic field (Skyrme number $N=1$). (a) Schematic of the tight‑focusing geometry. (b) Calculated intensity and phase (insets) distributions for the three Cartesian electric‑field components at the focal plane ($z=0$). (c) The major‑axis orientation of the polarization ellipses evaluated along the closed red circle indicated in (b). (d) Mapping of a family of spatial curves lying on the surface of a geometric M{\"o}bius snail into the physical focal volume. (e) The optical M{\"o}bius snail obtained by computing the major axis of the polarization ellipse at every sampled point along the curves shown in (d).}
	\label{fig2}
\end{figure}

As illustrated in Fig. \ref{fig2}(a), we consider a skyrmionic beam with a Skyrme number of $N=1$ (generated by superposing LG modes with indices $l_1=0$, $l_2=1$). This beam is tightly focused by a high-numerical-aperture (NA = 1.3) microscope objective, generating a complex 3D focal field. The resulting intensity distributions ($|E_x|^2$, $|E_y|^2$, and $|E_z|^2$) and phase profiles ($\Phi_x$, $\Phi_y$, and $\Phi_z$) for each Cartesian component of the focal field ($z=0$) are displayed in Fig. \ref{fig2}(b) (see Supporting Information for details). Consistent with earlier work \cite{Science-347-964-2015,PRL-117-013601-2016,Optica-7-1228-2020,NP-2014-8-23}, tracing the 3D orientation of the polarization ellipse along a closed loop surrounding the optical axis \revise{[e.g., the red circle indicated in Fig. 2(b), where the field components in circle have significant amplitude]}, reveals that the spatial configuration of the major axes constructs a M{\"o}bius strip, as depicted in Fig. \ref{fig2}(c). Within the complex-vector representation of the focal field, the major and minor axes of the polarization ellipse are correspondingly described by the vectors \cite{Science-347-964-2015}
\begin{equation}
	\begin{aligned}
		&\mathbf{A}=1/|\sqrt{\mathbf{E}\cdot \mathbf{E}}|{\rm Re}(\mathbf{E}^*\sqrt{\mathbf{E}\cdot \mathbf{E}})  \\
		&\mathbf{B}=1/|\sqrt{\mathbf{E}\cdot \mathbf{E}}|{\rm Im}(\mathbf{E}^*\sqrt{\mathbf{E}\cdot \mathbf{E}})
	\end{aligned}
\end{equation}
where $\mathbf{E}=(E_x,E_y,E_z)$ denotes the vectorial electric field in the focal region, ${\rm Re}(\cdot)$ and ${\rm Im}(\cdot)$ represent the real and imaginary parts, respectively.

\begin{figure*}[!htb]
	\centering
	\includegraphics [width=0.8\textwidth]{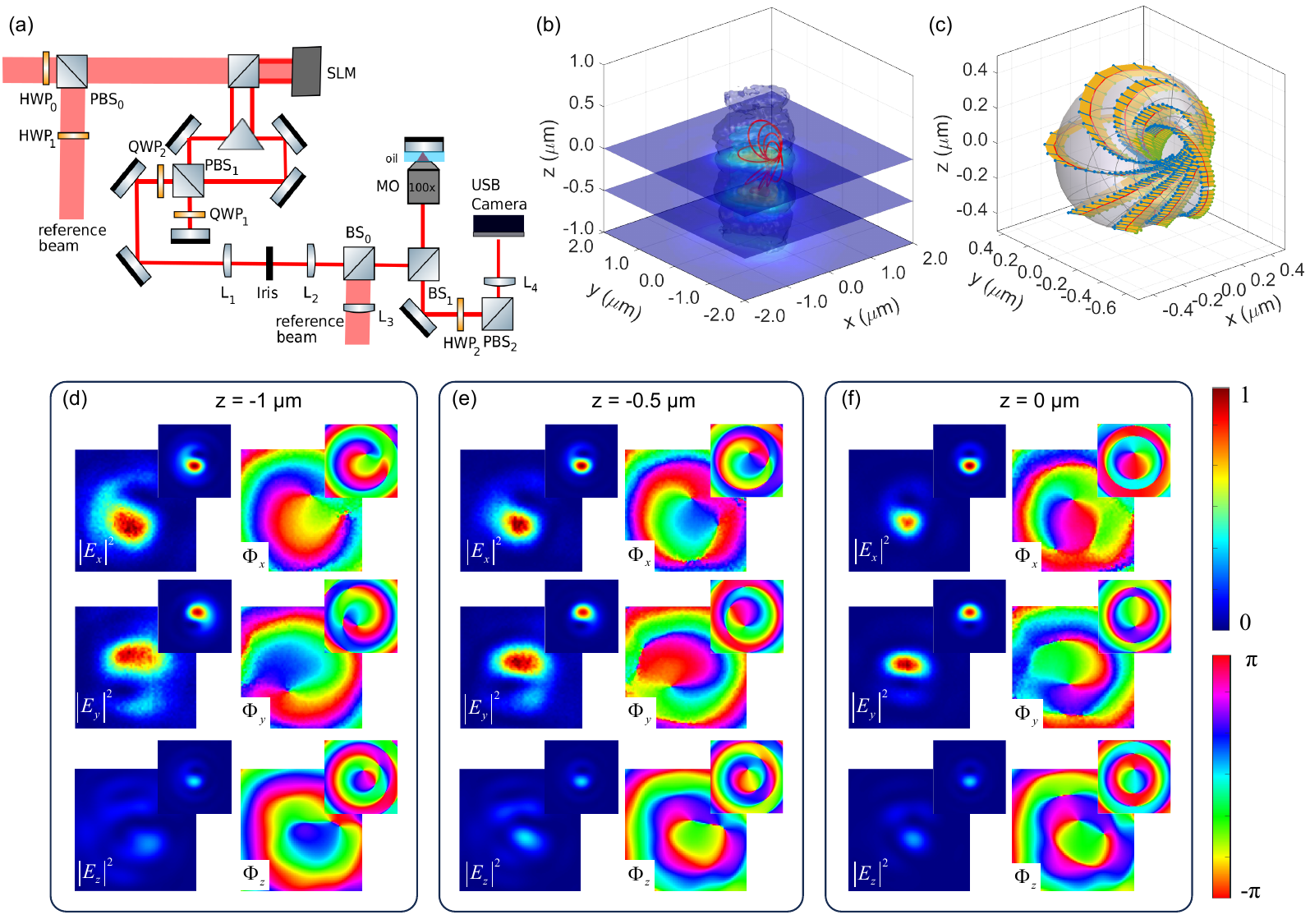}
	\caption{Experimentally observed optical M{\"o}bius snail in a tightly focused skyrmionic field (Skyrme number $N=1$). (a) Schematic of the experiment (not to scale). The focal length of $L1$, $L_2$, $L_3$ and $L_4$ are $f_1 =f_2=f_3=50\,$cm and $f_4=20\,$cm, respectively. (b) Measured total intensity distribution ($|E_x|^2+|E_y|^2+|E_z|^2$) in the 3D focal region, and red curves shows the mapping of a family of spatial curves lying on the surface of a geometric M{\"o}bius snail into the physical focal volume. (c) The 3D reconstruction of the optical M{\"o}bius snail. (d)-(f) Experimentally retrieved normalized intensity (left) and relative phase (right) distributions for the $E_x$,$E_y$ and $E_z$ field components at three selected longitudinal positions: z = -1 $\upmu$m, z = -0.5 $\upmu$m and z = 0 $\upmu$m (focal plane). The insets in the upper-right corners show the corresponding theoretical simulations.}
	\label{fig3}
\end{figure*}  

Inspired by the geometric evolution from a M{\"o}bius strip to the M{\"o}bius snail, we further explore the more intricate topology of polarization in the 3D vicinity of the focal volume. The resultant optical M{\"o}bius snail is presented in Figs. \ref{fig2}(d) and \ref{fig2}(e). To construct this entity, we first map a family of coordinate curves from the surface of a geometric M{\"o}bius snail into the physical space of the focal field via suitable translation and scaling [Fig. \ref{fig2}(d)]. These spatial curves sample a volume surrounding the focus, with the specific curve lying in the plane $z=0$ designed to coincide exactly with the closed loop used to generate the conventional M{\"o}bius strip in Fig. \ref{fig2}(b). At each sampled point along these curves, we calculate the local major axis $\mathbf{A}$ of the polarization ellipse using the calculated vector field $\mathbf{E}$. This procedure generates a series of distinct optical M{\"o}bius strips. Notably, these strips are not isolated, they are spatially interwoven within $\mathbb{R}^3$ and are intricately linked to the underlying scaffold of the geometric M{\"o}bius snail. Their collective evolution along this scaffold forms the coherent, volumetric structure we identify as an optical M{\"o}bius snail [Fig. \ref{fig2}(e)]. This construct thus provides a comprehensive 3D representation of the complex polarization topology inherent to a tightly focused skyrmionic field, moving beyond the planar description offered by a single M{\"o}bius strip.

To experimentally verify the predicted 3D topological structure, we directly reconstructed the complete electric field within the focal volume. This was achieved using a well-established phase-shifting interferometric technique \cite{PRA-17-064026-2022,JO-25-035602-2023,OL-48-3693-2023}, which allowed us to extract the hidden optical M{\"o}bius snail (the volumetric polarization topology) from the measured complex field. A simplified schematic of the experimental setup is shown in Fig. \ref{fig3}(a). The expanded CW laser at a wavelength ($\lambda$) of 1064\,nm is divided into reference and main beam by a half wave plate (HWP$_0$) and a polarizing beam splitter cube (PBS$_0$). The polarization of the reference beam is rotated by HWP$_1$ and then it is focused (with lens L$_3$) into the back aperture of the microscope objective (MO), collimating the reference. The fields are prepared with a spatial light modulator (SLM) illuminated by the main beam. The SLM screen is divided into two parts that independently encode the ${\rm LG}_{0,l_1}$ and ${\rm LG}_{0,l_2}$, since the SLM modulates only the horizontal polarization component, both fields have that polarization after been reflected by the SLM. The polarization of one of the beams is rotated to a vertical state (double pass through a quarter wave plate QWP$_1$) and both beams are combined with PBS$_1$. The paraxial vector beam propagates through a QWP$_2$ that sets the polarizations to $|L\rangle$ and $|R\rangle$ respectively. The vector beams are focused (lens L$_1$), spatially filtered (iris) and collimated (lens L$_2$). Finally, the vector beam ${\rm LG}_{0,l_1} |R\rangle + {\rm LG}_{0,l_2} |L\rangle$ is \revise{focused with the 
	oil-immersion microscope objective MO (100x, NA=1.3)}.

The tightly focused beam is retro-reflected by a mirror mounted into a piezo electric stage that controls its axial position and hence the axial cross-section  of the tightly focused beam that is imaged. The polarization of the imaged beam is controlled with HWP$_2$ and PBS$_2$, then lens L$_4$ projects the magnified image into a USB camera. 
The 3D focal field was probed by performing a fine axial scan with a piezoelectric stage over a range of $\pm1.025~ \upmu$m (~$\approx \pm1 ~\lambda$) centered on the nominal focal plane ($z=0$). The complex transverse field components, $E_x$ and $E_y$, were retrieved at each of the 41 sampled axial planes (separated by 50 nm) via a standard three-step phase-shifting interferometry (see Supporting Information). Finally, the longitudinal field component $E_z$ was calculated directly from the reconstructed transverse fields by enforcing the free-space divergence condition $\nabla \cdot \mathbf{E}=0$ \cite{PRL-135-033805-2025,APR-4-2300236-2023,AP-8-016003-2026}, ensuring a self-consistent, full-vectorial reconstruction of the focal field. 

\begin{figure*}[!htb]
	\centering
	\includegraphics [width=0.8\textwidth]{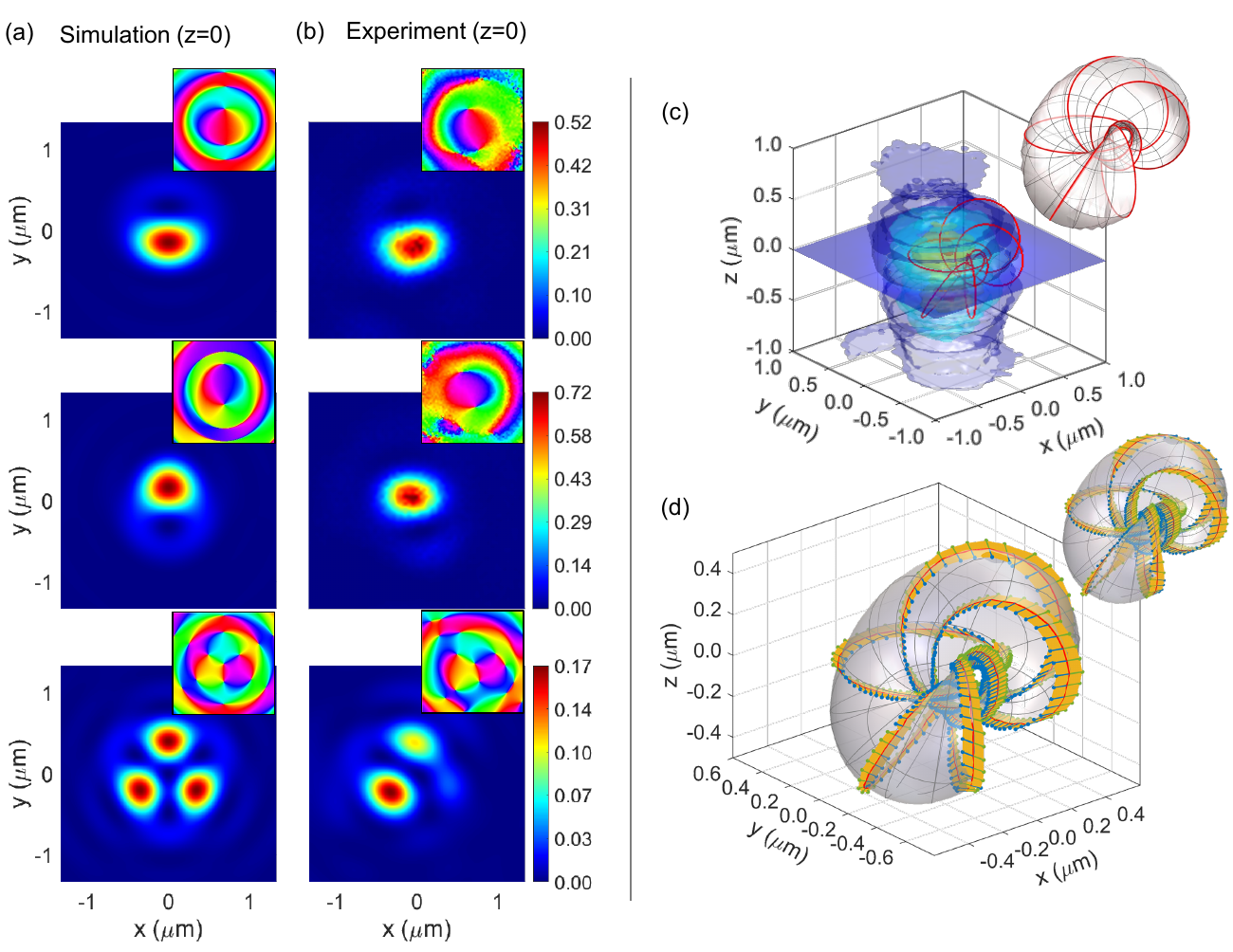}
	\caption{The experimentally observed high-order optical M{\"o}bius snail in tightly focused field when $N=-1$. (a) and (b) Simulated and experimentally measured intensity and phase distributions of the three field components at the focal plane ($z=0$). (c) Mapping from a series of curves on a higher-order M{\"o}bius snail surface to the measured 3D focal field. (d) Reconstruction of the optical M{\"o}bius snail based on the experimentally observed major-axis orientations of polarization ellipses within the 3D field. Top-right insets show the corresponding simulated results. }
	\label{fig4}
\end{figure*} 

The experimentally reconstructed total intensity distribution in the 3D focal region is presented in Fig. \ref{fig3}(b). To unveil the underlying topological structure, we computed the major axis $\mathbf{A}$ of the polarization ellipse at discrete locations along a set of predefined spatial curves within this volume (marked by the red curves in Fig. \ref{fig3}(b)). The resulting three-dimensional arrangement of polarization axes faithfully reproduces the characteristic twisted and interwoven geometry of an optical M{\"o}bius snail, as directly visualized in Fig. \ref{fig3}(c). This reconstruction provides \revise{experimental confirmation} of the volumetric polarization topology theorized in the preceding sections. To further validate the fidelity of our field retrieval, Figs. \ref{fig3}(d)–\ref{fig3}(f) display the normalized intensity and relative phase distributions for all three Cartesian field components ($E_x$, $E_y$ and $E_z$) on representative transverse planes at distinct longitudinal positions. The experimental data exhibit excellent agreement with the corresponding theoretical simulations (shown in the insets). Observed minor deviations in the intensity profiles and phase maps can be ascribed to typical experimental limitations, including wavefront aberrations and slight misalignment in the experimental setup. For a consistent visual comparison, the intensity of each field component is normalized to the total intensity within the respective transverse plane.

The topology of the optical M{\"o}bius snail is not fixed but can be \revise{controlled} by tailoring the parameters of the incident skyrmionic beam. A direct demonstration of such control is achieved by inverting the Skyrme number of the input field. Specifically, swapping the orbital angular momentum indices between the two circular polarization components in Eq. \ref{eq1} (i.e., setting $l_1=1$ and $l_2=0$), reverses the sign of the Skyrme number from $N=1$ to $N=-1$. Under tight focusing, this inversion of the topological charge fundamentally alters the 3D polarization texture: the winding structure reorganizes correspondingly. This transformation yields a distinct higher-order optical M{\"o}bius snail compared to the $N=1$ case (see Supporting Information for details). Such parameter-dependent reconfiguration illustrates the versatility and programmability of the optical M{\"o}bius snail in 3D structured light.

Figures \ref{fig4}(a) and \ref{fig4}(b) present the simulated and experimentally measured intensity and phase distributions (insets) for the three electric-field components ($E_x, E_y, E_z$) at the focal plane ($z=0$) in the $N=-1$ case. Compared with the single-twist case shown in Fig. \ref{fig2}(b), this represents a multi‑twist polarization ribbon topology confined to a plane \cite{Science-347-964-2015, NJP-2019-21-053020}. When extended into three dimensions, its evolution gives rise to higher‑order multitwist Möbius snails. While minor discrepancies between the experimental results and theoretical simulations are present (likely due to wavefront distortions or alignment inaccuracies), these imperfections do not obscure the identification of the underlying higher-order topological features. These results correspond essentially to the type of optical Möbius strip reported previously, in which a single closed loop in the transverse plane gives rise to a twisted ribbon structure defined by the major axis of the polarization ellipse \cite{Science-347-964-2015}. Despite differences in experimental parameters (e.g., wavelength and NA), our results at $z=0$ reproduce this fundamental geometry, thereby confirming consistency with earlier observations. However, beyond this planar cross-section, our work reveals a significantly richer 3D topology: by extending the analysis to a family of loops displaced from the focal plane, we uncover a ensemble of Möbius strips that collectively form the “Möbius snail” -- a continuous 3D topological texture that has not been reported before. To explicitly reveal this structure, a family of spatial curves defined on the surface of a higher-order Möbius snail is mapped into the reconstructed 3D focal field, as illustrated in Fig. \ref{fig4}(c). By calculating the major axis $\mathbf{A}$ of the polarization ellipse at every sampled point along these spatial trajectories, we obtain a twisted bundle of polarization axes that interweave coherently in three dimensions. The resulting higher-order optical M{\"o}bius snail is displayed in Fig. \ref{fig4}(d). The experimentally reconstructed geometry closely matches the theoretically predicted structure shown in the inset, confirming that the sign reversal of the Skyrme number directly changes the 3D polarization topology. This correspondence demonstrates that the global topology of a tightly focused field can be controllably engineered via the topological charge of the incident beam, and establishes interferometric reconstruction as a reliable technique for visualizing such complex 3D polarization textures.

In summary, our work introduces and generates the optical M{\"o}bius snail--a 3D topological structure in a tightly focused light field, and it complements existing 3D structurd light such as hopfions, torons, and knots. We establish a general mapping that directly links the geometric M{\"o}bius snail to the global winding of the polarization ellipse’s major axis in a tightly focused vectorial field. Using full-vectorial interferometric tomography, we visualized this 3D topology and further demonstrated its \revise{switchable topology by changing} the sign of the incident Skyrme number (from $N=1$ to $N=-1$) of skyrmionic beam. \revise{This provides a simple method to generate and switch between two distinct polarization topologies.}

Furthermore, from a mathematical perspective, the M{\"o}bius snail is isomorphic to the Clifford torus and the Klein bottle (two fundamental topological surfaces) \cite{PP-32-102503-2025}. This equivalence suggests that our approach can be extended to realize other complex topological geometries in light, such as toroidal or higher-genus structures, thereby enriching the family of 3D topological photonic states. The resulting M{\"o}bius snail inherit \revise{topological structures}, making them suitable as a possible platform for exploring topological invariants in higher dimensions and studying their behavior under perturbations \cite{NC-16-3001-2025,NC-2026}. Extending this concept beyond optics, we anticipate that analogous snail-like topological textures \revise{may be} realizable in other wave systems, including acoustics, hydrodynamics, liquid crystals, magnetic materials, and atomic Bose–Einstein condensates.\\

\noindent {\textbf{Funding.}} We acknowledge the support of the National Natural Science Foundation of China (12504353,12374308); Natural Science Foundation of Guangdong Province (2025A1515010738); Joint Laboratory of Guangdong, Hong Kong, and Macao Universities, Guangdong Province (2022LSYS006); National Key R\& D Program of China (2022YFA1404800); STU Scientific Research Initiation Grant (NTF24020T); Singapore Ministry of Education (MOE) AcRF Tier 1 grants (RG157/23 \& RT11/23); Singapore Agency for Science, Technology and Research (A*STAR) (M24N7c0080 \& H25-MRO3489), Nanyang Assistant Professorship Start Up grant; CIC UNAM; PAPIIT UNAM (IG101425), and Secihti (I-657).  \\

\noindent {\textbf{Acknowledgments.}} PAQS thanks C. Mojica-Casique for technical support and J. Rangel for machine shop work.\\

\noindent { \textbf{Conflict of Interest.}} The authors declare no conflicts of interest.\\

\noindent {\textbf{Data Availability Statement.}} The data that support the findings of this study are available from the corresponding author upon reasonable request.\\

\end{document}